\documentclass[%
 reprint,
 amsmath,amssymb,
prl,
]{revtex4-1}

\usepackage{graphicx}
\usepackage{dcolumn}
\usepackage{bm}

\begin{document}

\preprint{APS/123-QED}
\title{Surface solitons in $\cal{PT}$ symmetric potentials}
\author{Huagang Li,$^{1,2}$, Zhiwei Shi,$^3$ Xiujuan jiang,$^3$ Xing Zhu,$^1$ and Tianshu Lai,$^{1,}$}
\thanks{E-mail: stslts@mail.sysu.edu.cn (Tianshu Lai)}
\affiliation{$^1$State-Key Laboratory of Optoelectronic Materials and Technologies, School of Physics and Engineering,
Sun Yat-Sen University, Guangzhou 510275, China\\
$^2$Department of Physics, Guangdong University of Education , Guangzhou 510303, China\\
$^3$School of Information Engineering, Guangdong University of Technology, Guangzhou 510006, China}
\date{\today}
\begin{abstract}
We investigate light beam propagation along the interface between linear and nonlinear media
with parity-time ($\cal{PT}$) symmetry, and derive an equation governing the beam propagation. A novel class of two-dimensional $\cal{PT}$ surface solitons are found analytically and numerically, and checked to be stable over a wide range of $\cal{PT}$ potential structures and parameters. These surface solitons do not require a power threshold. The transverse power flow across beam is examined within the solitons and found to be caused by the nontrivial phase structure of the solitons. $\cal{PT}$ surface solitons are possibly observed experimentally in photorefractive crystals.

\begin{description}
\item[PACS numbers]
03.65.Ge, 11.30.Er, 42.65.Sf, 42.65.Tg
\end{description}
\end{abstract}

\pacs{Valid PACS appear here}
\maketitle

Surface solitons were found to be able to propagate along the interface between two media with real index of refraction,
and showed many interesting properties
which had no analogues in single medium alone~\cite{1-0,2-0,3-0}.
It was also reported that surface solitons could exist at the interface between
two media where at least one of them is nonlinear.
Theoretical and experimental studies on the surface solitons have been reported in two-dimensional domain
where many interesting phenomena associated with nonlinear surface waves were predicted and observed~\cite{2-2,2-3,2-4,2-5}

On the other hand, Bender and Boettcher found a non-Hermitian Hamiltonian might have an entirely real eigenvalue spectrum~\cite{7-0}
when potential field met parity-time ($\cal{PT}$) symmetry, which implied eigenstate might exist in $\cal{PT}$ symmetric complex potential field.
In optics, complex potentials mean a gain and loss contained in optical media.
Consequently, the solitons in $\cal{PT}$ symmetric optical potentials were explored extensively in theory and experiments thereafter.
Mussilimani studied $\cal{PT}$ solitons in one-dimensional (1D) and two-dimensional (2D) media~\cite{8-0,9-0}.
The basic properties of Floquet-Bloch modes in $\cal{PT}$ symmetric optical lattices were examined in detail~\cite{10-0}.
In 2010, optical $\cal{PT}$ symmetry was observed experimentally in an optical coupled system involving a complex index potential~\cite{11-0}.
$\cal{PT}$ symmetry breaking was also demonstrated in complex optical potentials~\cite{12-0,jia-s}.
The effect of localized modes in lattices of size N with exponentially fragile $\cal{PT}$ symmetry was also studied~\cite{13-0}.
The light transmission in large-scale temporal $\cal{PT}$ symmetric lattices was studied~\cite{14-0}.
It was also found that gray solitons might exist and be stabilized in $\cal{PT}$ symmetric potentials~\cite{15-0}.
However, all investigations mentioned above dealt with the $\cal{PT}$ symmetry and $\cal{PT}$ solitons only within one complex index medium.
The surface solitons at the interface between two complex index media, a generalized surface solitons, have not been explored.
It is still unknown if such surface solitons may exist and stabilize.

In this letter, we explore such surface solitons in two complex index media composed of self-focusing and linear media,
and find that the 2D surface solitons may exist and stabilize at 2D interface
when the optical potential of the complex index self-focusing medium meets $\cal{PT}$ symmetry and the interface is oriented in a particular direction.
We gain the 2D $\cal{PT}$ surface solitons, and test their stability numerically.

We express the optical field as $E(x,y,z,t)=E(x,y)e^{i(\Gamma z-\omega t)}$, the refractive index in nonlinear and linear media
as $n_{01}+n_{r1}(x,y)+i n_{i1}(x,y)+n_2|E|^2$ and $n_{02}+n_{r2}(x,y)+i n_{i2}(x,y)$, respectively, where $\omega$ is the frequency, $\Gamma$ is the wave vector,
$n_{01}$ and $n_{02}$ are the background refractive indices, $n_{r1,r2}(x,y)$ is the real index profile and $n_{i1,i2}(x,y)$
the imaginary part (gain or loss) of $\cal{PT}$ symmetric potentials, and $n_2$ is nonlinear Kerr coefficient~\cite{8-0,9-0}.
Substituting the field and the refractive index into the Helmholtz equation, we get,
\begin{eqnarray}
 i2\Gamma\frac{d E}{d z}+\frac{\partial^2 E}{\partial x^2}+\frac{\partial^2 E}{\partial y^2}+(k_0^2n_{01}^2-\Gamma^2)E \nonumber\\
   +2k_0^2[n_{01}(n_{r1}+i n_{i1})+n_{01}n_2|E|^2]E=0,
  \label{eq:one}
\end{eqnarray}
in the nonlinear medium when the conditions, $k_0=\omega/c$, $n_{r1,i1}<<n_{01}$ and $n_2|E|^2<<n_{01}$, were met, and
\begin{eqnarray}
i2\Gamma\frac{d E}{d z}+\frac{\partial^2 E}{\partial x^2}+\frac{\partial^2 E}{\partial y^2}+(k_0^2n_{02}^2-\Gamma^2)E \nonumber\\
 +2k_0^2n_{02}(n_{r2}+i n_{i2})E=0,
 \label{eq:two}
\end{eqnarray}
in the linear medium as $n_{r2,i2}<<n_{02}$.

Introducing the scaled dimensionless quantities as $\zeta=z/(2\Gamma w_0^2)$, $\xi=x/w_0$, $\eta=y/w_0$,
$q=n_{01}^{\frac{1}{2}}/[k_0n_{01} w_0 (2n_2)^{\frac{1}{2}}]E$,
$R=2n_{01}k_0^2w_0^2(n_{r1}+i n_{i1})$ and $R=2n_{02}k_0^2w_0^2(n_{r2}+i n_{i2})$, respectively in nonlinear and linear media,
 where $w_0$ is an arbitrary scaling factor,
Eqs.(\ref{eq:one}) and (\ref{eq:two}) can be changed as,
\begin{equation}
 i\frac{d q}{d \zeta}+\nabla_\perp^2q+Rq+G_1q+|q|^2q=0,
\label{eq:three}
 \end{equation}
 \begin{equation}
 i\frac{d q}{d \zeta}+\nabla_\perp^2q+Rq+G_2q=0,
 \label{eq:four}
\end{equation}
where $G_1=w_0^2[(k_0n_{01})^2-\Gamma^2]$, $G_2=w_0^2[(k_0n_{02})^2-\Gamma^2]$ and $\nabla_\perp^2=\partial^2/ \partial \xi^2 +\partial^2 /\partial \eta^2$.
Setting $q=q'e^{iG_1\zeta}$, Eqs.(\ref{eq:three}) and (\ref{eq:four}) can be changed as follows,
\begin{equation}
 i\frac{d q'}{d \zeta}+\nabla_\perp^2q'+Rq'+|q'|^2q'=0,
\label{eq:seven}
 \end{equation}
 \begin{equation}
 i\frac{d q'}{d \zeta}+\nabla_\perp^2q'+Rq'+\Delta Gq'=0,
 \label{eq:eight}
\end{equation}
where $\Delta G=G_2-G_1=w_0^2k_0^2(n_{02}^2-n_{01}^2)$. Based on the step function $U(x,y)$ which is equal to 1 or 0, respectively in nonlinear and linear media, Eqs.(\ref{eq:seven}) and (\ref{eq:eight}) can be integrated
into one equation as follows,
\begin{equation}
 i\frac{d q'}{d \zeta}+\nabla_\perp^2q'+Rq'+(1-U)\Delta Gq'+U|q'|^2q'=0,
 \label{eq:jia_1}
\end{equation}

Assuming a stationary soliton solution to be in the form $q'(\xi,\eta,\zeta)=u(\xi,\eta)e^{ib\zeta}$~\cite{16-0},
where $u$ is a complex function and $b$ is the propagation constant,
Eq.(\ref{eq:jia_1}) can now be deduced to,
\begin{equation}
 \nabla_\perp^2u+[R+(1-U)\Delta G]u+U|u|^2u-bu=0.
 \label{eq:five}
 \end{equation}

$R+(1-U)\Delta G$ is a $\cal{PT}$ symmetric potential because $R$ has $\cal{PT}$ symmetry. As a result, Eq.(\ref{eq:five}) describes a stationary $\cal{PT}$ soliton~\cite{8-0} in the $\cal{PT}$ potential, $R+(1-U)\Delta G$.

To seek for possible analytical stationary soliton solutions to  Eq.(\ref{eq:five}),  a special $\cal{PT}$ symmetric potential, $R=V+iW$, is constructed, which is expressed by,
\begin{eqnarray}
&V=(4+\frac{W_0^2}{18})\sec h^2(\xi+\eta)+4 \sec h^2(\xi-\eta) \nonumber\\
&-[1+e^{\alpha(\xi-\eta)}]^{-2\alpha} \sec h^2(\xi+\eta)\sec h^2(\xi-\eta),
\label{eq:j_1}
\end{eqnarray}
\begin{equation}
W=W_0\sec h(\xi+\eta)\tanh(\xi+\eta),
\label{eq:j_2}
\end{equation}
\begin{eqnarray}
&(1-U)\Delta G=-\frac{2\alpha^4 +2\alpha^3}{[1+e^{\alpha(\eta-\xi)}]^2}
+\frac{2\alpha^3-4\alpha^2\tanh(\xi-\eta)}{1+e^{\alpha(\eta-\xi)}},
\label{eq:j_3}
\end{eqnarray}
where the step function $U(\xi,\eta)$ is approximated by the function, $1/[1+e^{\alpha(\xi-\eta)}]$,
to guarantee its differentiability.
Meanwhile, this approximate step function also implies that the interface between nonlinear and linear media
coincides with the line $\xi-\eta=0$.
$W_{0}$ is the amplitude of the imaginary part of the $\cal{PT}$ symmetric potentials,
$\alpha$ is a positive factor which describes the degree of proximity between $1/[1+e^{\alpha(\xi-\eta)}]$
and $U(\xi,\eta)$.

For the $\cal{PT}$ symmetric potential described by Eqs.(\ref{eq:j_1})-(\ref{eq:j_3}), it is possible to realize it in a photorefractive crystal by means of appropriate light pumping through specially designed masks, as shown in~\cite{11-0}, and an analytical stationary soliton solution to Eq.(\ref{eq:five}) is found as follows,
\begin{eqnarray}
u=\frac{\sec h(\xi+\eta) \sec h(\xi-\eta)}{[1+e^{\alpha(\xi-\eta)}]^\alpha} e^{i \frac{W_0}{6} \arctan[\sinh(\xi+\eta)]},
\label{eq:j_4}
\end{eqnarray}
with the propagation constant $b=4$.

To shed light on the formation of the $\cal{PT}$ surface solution,
we examine the transverse power-flow density across the beam, which can be described by Poynting vector~\cite{8-0},
$\vec{S}=(i/2)(u\nabla u^*-u^*\nabla u)$, where $\nabla=\partial/\partial \xi+ i \partial/\partial \eta$,
and intuitively indicates the energy exchange.
For the soliton in Eq.(12), we get,
\begin{eqnarray}
&\vec{S}=\frac{  (1+i)W_0\sec h^2(\xi-\eta) \sec h^3(\xi+\eta)}{6[1+e^{\alpha(\xi-\eta)}]^{2\alpha}}.
\end{eqnarray}
It is interesting to note that the energy-flow vector $\vec{S}$ is always parallel to the interface and along the positive direction of $\xi$ axis, implying that the power always flows
from the gain toward the loss region, and no energy exchange between nonlinear and linear media.
Because $1/[1+e^{\alpha(\xi-\eta)}]^{2\alpha}$ is similar to the step function $U(\xi,\eta)$, the power-flow density is mainly concentrated in the nonlinear medium.
\begin{figure}[htb]
\centerline{\includegraphics[width=10cm]{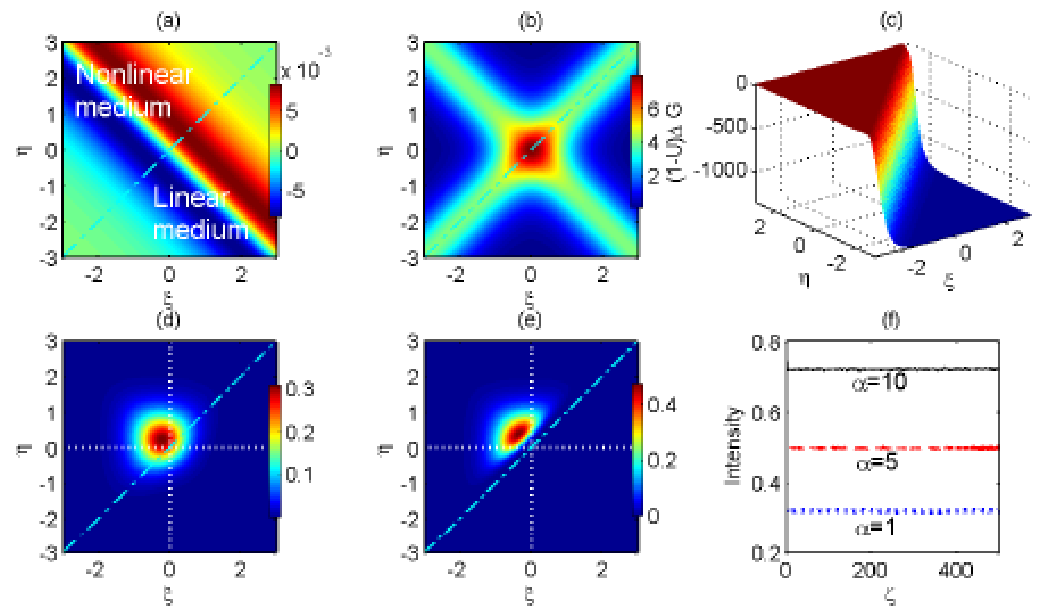}}
\caption{\label{fig:one} (a) The imaginary and the real (b) parts of the potential $R$ with $\alpha=5$.
(c) The profile of $(1-U)\Delta G$ based on Eq.(\ref{eq:j_3}). (d) and (e) The intensity distribution of the analytical surface solitons with $\alpha=1$ and $5$, respectively.
(f) The maximum intensity of the propagating surface solitons versus the propagation distance with $5\%$ noise for different $\alpha$. The dot-dash line displays the interface between the two media.
The other parameter is $W_0=0.1$.}
\end{figure}

The imaginary and real parts of the potential $R$, and $(1-U)\Delta G$ based on Eq.(\ref{eq:j_3}), respectively,
are displayed in Fig.~\ref{fig:one} (a), (b) and (c) with $\alpha=5$ and $W_0=0.1$.
The intensity profiles of the analytical surface solitons are plotted in Fig.~\ref{fig:one} (d) and (e) with $\alpha=1$ and $5$, respectively.
It is clearly seen that the center of the soliton profile does not locate at the interface when $\alpha$ is bigger,
but is slightly away from the interface into the nonlinear medium along the odd symmetric axis ($\xi+\eta=0$) of the $\cal{PT}$ symmetric potentials
so that only a small fraction of the soliton power is existed within the linear medium, as shown in Fig.~\ref{fig:one} (e),
which is the result of the initial net transverse power flow from the linear to nonlinear media.
The intensity profile of the soliton is symmetrical in both gain and loss regions.
Such surface solitons can be formed and stabilized because the net gain and loss, respectively in the gain and loss regions
is zero due to transverse power flow from gain to loss regions, while the power flow arises from the nontrivial phase structure of the surface solitons~\cite{8-0}, as the soliton solution shows in Eq. (\ref{eq:j_4}).
We have checked the robustness of the analytical surface soliton by simulating the soliton propagation
with 5\% random noise imposed on both amplitude and phase.
The maximum intensity of the soliton in propagation is plotted in Fig.~\ref{fig:one} (f),
showing the stability of the soliton for different $\alpha$.
One can see that the stability of the soliton is very well.
\begin{figure}[htb]
\centerline{\includegraphics[width=10cm]{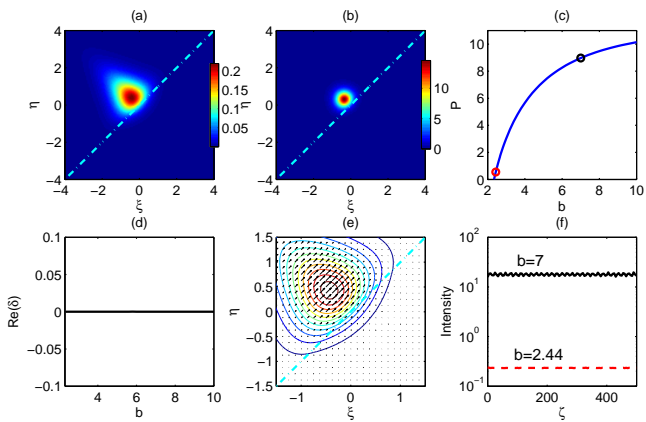}}
\caption{\label{fig:two}  (color online) The intensity profile of the solitons with the propagation constant $b=2.44$ (a) and $7$ (b).
(c) The power $P$ of the surface soliton versus $b$. The red and black filled circles correspond to the solitons in (a) and (b), respectively.
(d) The real part of the perturbation growth rate $\delta$ versus $b$.
(e) Transverse power flow of the soliton in (a) where the contour plot in the background represents
the its intensity distribution.
(f) The maximum intensity of the surface solitons in (a) and (b) propagate with $5\%$ noise.
The other parameters include $W_0=0.1$, $\Delta G=-1$. }
\end{figure}

From Fig.~\ref{fig:one} (d) and (e), one can see that the power of the surface solitons confined in the linear medium is very small or almost zero
when $|\Delta G|$ or $\alpha$ is bigger.
Accordingly, we can draw a conclusion that $\cal{PT}$ surface solitons can exist and stabilize
even if $\cal{PT}$ symmetric potential exists only in nonlinear medium regardless of $\cal{PT}$ symmetry of the linear medium because only very small portion of the soliton power is in linear medium. To check the conclusion, the $\cal{PT}$ potential in Eqs.(\ref{eq:j_1})-(\ref{eq:j_3}) is modified as $V=U(\xi,\eta)[(4+\frac{W_0^2}{18})\sec h^2(\xi+\eta)+4 \sec h^2(\xi-\eta)-\sec h^2(\xi+\eta)\sec h^2(\xi-\eta)]$,
and $W=W_0U(\xi,\eta)\sec h(\xi+\eta)\tanh(\xi+\eta)$, that is to ignore the $\cal{PT}$ potential in the linear medium.
Numerical soliton solutions to Eq.(\ref{eq:five}) under the modified $\cal{PT}$ potential are obtained using the spectral renormalization method~\cite{17-0}, and are displayed in Fig.~\ref{fig:two} (a) and (b) with $b=2.44$ and $b=7$, respectively, for $\Delta G=-1$ and $W_0=0.1$.
The spot size of the solitons is obviously affected by the propagation constant $b$, as shown in Fig.~\ref{fig:two}(a) and (b).
Furthermore, the intensity of the solitons increases with $b$.
The power of solitons, $P=\int^\infty_{-\infty}\int^\infty_{-\infty}|u|^2d\xi d\eta$, is plotted in Fig.~\ref{fig:two} (c)
as a function of $b$, and is obviously lower than the self-focusing critical power ($P_{cr}\approx11.70$)~\cite{19-0}.
As a consequence, the formation of the $\cal{PT}$ surface solitons is no requirement of critical power threshold.

The stability of the surface solitons is also checked with the linear stability analysis method.
It is assumed that $q'(\xi,\eta)=u(\xi,\eta)e^{ib\zeta}+\epsilon[F(\xi,\eta)e^{\delta \zeta}+G^*(\xi,\eta)e^{\delta^*\zeta}]e^{ib\zeta}$~\cite{8-0,18-0},
where $\epsilon\ll1$, $F$ and $G$ are the perturbation eigenfunctions and $\delta$ is the growth rate of the perturbation.
Substituting $q'(\xi,\eta)$ into Eq.(\ref{eq:jia_1}) and linearizing the equation, we gain,
\begin{eqnarray}
&\delta F=i[(\nabla_\perp^2+R+\psi+2U|u|^2)F+Uu^2 G],\\
&\delta G=-i[(\nabla_\perp^2+R^*u+\psi+2U|u|^2)G+Uu^{*2} F],
\end{eqnarray}
where $\psi(\xi,\eta)=[1-U(\xi,\eta)]\Delta G-b$.

It is well known that the soliton is linearly unstable if the real part of growth rate $\delta$ is positive. Otherwise, it is linearly stable~\cite{18-0}.
Fig.~\ref{fig:two} (d) shows the Re($\delta$) versus $b$. One can see the real part of $\delta$ is not positive as $b\geq2.35$.
As a result, the surface solitons are always stable until the power of the solitons is close to the self-focusing critical power $P_{cr}$.
Numerical results also show that the surface solitons do not exist as $b<2.35$.
The transverse power flow (Poynting vector) is mapped by arrows in Fig.~\ref{fig:two} (e),
which intuitively indicates the energy exchange among gain or loss domains.
The robustness of the surface solitons has also been checked with 5\% random noise.
The maximum intensity of the solitons in propagation is plotted in Fig.~\ref{fig:two} (f),
showing the stability of the solitons for different $b$.
\begin{figure}[htb]
\centerline{\includegraphics[width=10cm]{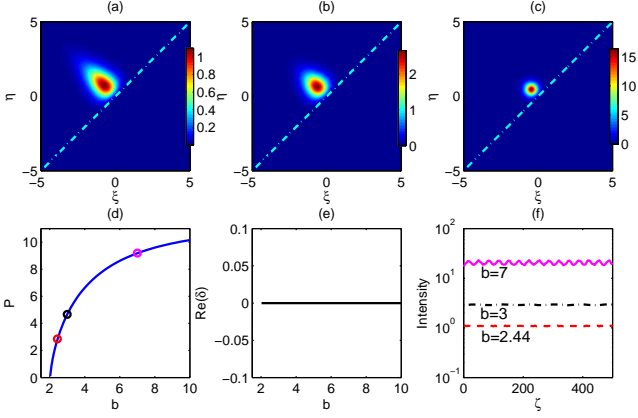}}
\caption{\label{fig:three} (color online) The intensity profile of the solitons with $b=2.44$ (a), $3$ (b) and  $7$ (c).
The dot-dash line displays the interface between the two media.
(d) The power $P$ of the surface solitons versus $b$.
The  red, black, and pink filled circles correspond to the solitons in (a), (b) and (c), respectively.
(e) The real part of the perturbation growth rate $\delta$ versus $b$.
(f) The maximum intensity of the surface solitons propagates with $5\%$ noise.
The  red, black and pink curves correspond to the solitons in (a), (b) and (c), respectively.
The other parameters are $W_0=0.1$, $\Delta G=-50$. }
\end{figure}

Next, setting $\Delta G=-50$ (corresponding to a larger refractive difference $n_{02}^2-n_{01}^2$ in the modified potential), numerical surface solitons are calculated and plotted in Fig.~\ref{fig:three} (a), (b) and (c) for different propagation constant $b$.
One can see that the power of the surface soliton confined in the linear medium is obviously less than one in Fig.~\ref{fig:two} (a) and (b) where $|\Delta G|$ is smaller than here.
The total power of the solitons can also be found still lower than the self-focusing critical power, as Fig.~\ref{fig:three} (d) shows,
and hence no power threshold required for the formation of surface solitons.
Fig.~\ref{fig:three} (e) shows the stable region of the surface solitons, which is similar to one in Fig.~\ref{fig:two} (d).
The maximum intensity of the solitons in propagation is plotted in Fig.~\ref{fig:three} (f), also showing that the soliton is stable,
and revealing the agreement with the analysis of linear stability in Fig.~\ref{fig:three} (e).
\begin{figure}[htb]
\centerline{\includegraphics[width=10cm]{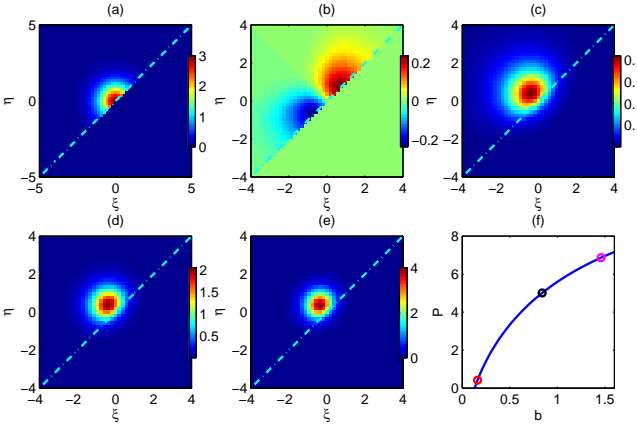}}
\caption{\label{fig:four} (color online) (a) The real and (b) the imaginary parts of the modified Scarff II potential $R$.
The intensity profile of the solitons with $b=0.16$ (c), $0.84$ (d) and $1.46$ (e), respectively.
(f) The power $P$ of the surface solitons versus propagation constant $b$.
The  red, black and pink filled circles correspond to the solitons in (c), (d) and (e), respectively.
The dot-dash line displays the interface between the two media. The other parameters are $V_0=3$, $W_0=0.3$, $\Delta G=-1$. }
\end{figure}

To explore the generalization of the $\cal{PT}$ surface soliton, its existence and stability are investigated in another kind of potentials, a two-dimensional Scarff II potential~\cite{8-0}, which is expressed by
$R= U(\xi,\eta)\{V_0\sec h^2(\sqrt{\xi^2+\eta^2})+iW_0 \sec h(\sqrt{\xi^2+\eta^2})[\tanh(\xi)+\tanh(\eta)]\}$,
where $U(\xi, \eta)$ is still a step function as depicted before.
The numerical $\cal{PT}$ surface soliton solution to Eq.(\ref{eq:five}) is obtained under the two-dimensional Scarff II potential.
The real and imaginary parts of the potential $R$ are displayed in Fig.~\ref{fig:four} (a) and (b), respectively, while
the typical surface solitons are plotted in Fig.~\ref{fig:four} (c), (d) and (e), respectively with $b=0.16$, $0.84$ and $1.46$ for $V_0=3$, $W_0=0.3$ and $\Delta G=-1$.
The power of the soliton is plotted in Fig.~\ref{fig:four} (f)
as a function of $b$, and is again shown lower than the self-focusing critical power $P_{cr}$, which is similar to the result in above other potentials.

\begin{figure}[htb]
\centerline{\includegraphics[width=10cm]{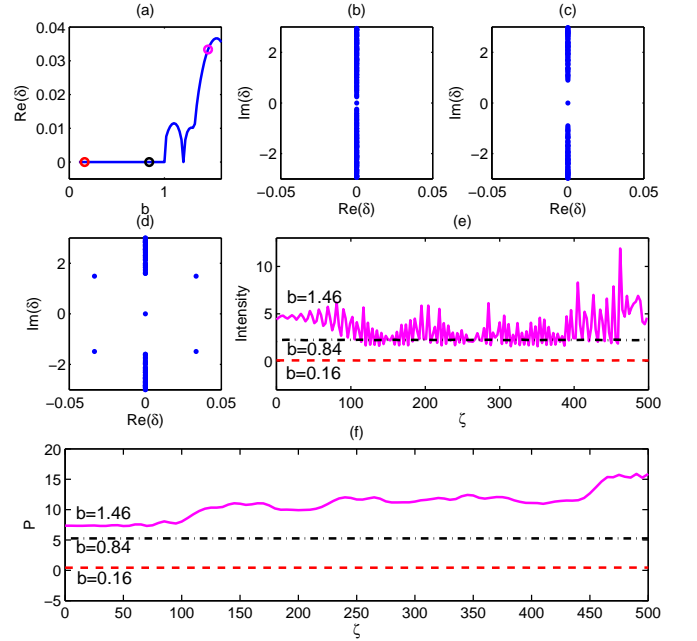}}
\caption{\label{fig:five}
(color online) (a) The real part of the perturbation growth rate $\delta$ versus $b$.
The red, black and pink filled circles correspond to the solitons in Fig.~\ref{fig:four} (c), (d) and (e), respectively.
Linear stability spectra of the soliton with $b=0.16$ (b), $0.84$ (c) and $1.46$ (d).
(e) The maximum intensity and (f) the power of the surface solitons in propagation with $5\%$ noise corresponding to
Fig.~\ref{fig:four} (c), (d) and (e), respectively.}
\end{figure}

The stability of surface solitons may be measured by the real part of perturbation growth rate. The real parts of the growth rate versus $b$ are
plotted in Fig.~\ref{fig:five} (a). One can find that the stability of the solitons maintains only in a small range of $b$, which is much different from the case in the other
potentials before where the soliton can stabilize in a wide range of $b$, as shown in Fig.~\ref{fig:two} (d) and Fig.~\ref{fig:three} (e). Fig.~\ref{fig:five} (a) shows the soliton in Fig.~\ref{fig:four} (e) ($b=1.46$) is unstable.
Fig.~\ref{fig:five} (b), (c) and (d) show the linear stability spectra of the solitons , respectively, corresponding to the solitons in
Fig.~\ref{fig:four} (c), (d) and (e). Fig.~\ref{fig:five} (d) still shows the unstability of the soliton in Fig.~\ref{fig:four} (e).
The maximum intensity and power of the solitons in propagation are displayed in Fig.~\ref{fig:five} (e) and (f), respectively. They all show that the soliton with $b=1.46$ is
unstable due to the oscillation of the intensity and power in propagation till divergence eventually.

In summary, a novel class of two-dimensional surface solitons is reported. It exists at the interface between the linear and nonlinear complex refractive index media with $\cal{PT}$ symmetry. The soliton equation governing the evolution of the surface soliton is derived. Analytical and numerical solition solutions are explored over a wide range of $\cal{PT}$ potential structures and parameters. It is found that the surface soliton can stabilize over a wide range of potential structures and parameters, and even can exist and stabilize at the interface where nonlinear medium alone meets $\cal{PT}$ symmetry but linear medium does not. One-dimensional $\cal{PT}$ surface soliton can not exist, which is distinct from the case in single $\cal{PT}$ symmetric potential where one-dimensional $\cal{PT}$ soliton can exist and stabilize within the $\cal{PT}$ potential. The cause that one-dimensional $\cal{PT}$ surface soliton can not exist is that the gain and loss regions are alone distributed in one of the linear and nonlinear media, while the interface between the two media prevents free power flow from the gain to loss domains. The nontrivial phase structure of $\cal{PT}$ surface solitons plays a key role in maintaining stability of the $\cal{PT}$ surface solitons. It is the nontrivial phase structure that leads to the transverse power flow from gain to loss domains to keep energy conservation in the propagation of the $\cal{PT}$ surface solitons along the interface.
Furthermore, these surface solitons do not exhibit a requirement of power threshold.

This research is partially supported by the National Natural Science Foundation of China under grant Nos.11204043, 11274399 and 61078027, National Basic Research Program of China under grant Nos. 2010CB923200 and 2013CB922403, and doctoral specialized fund of MOE of China under grant No. 20090171110005).

\end{document}